# An In-Depth Analysis of Ride-Hailing Travel Using a Large-scale Trip-Based Dataset

**Jianhe Du, Ph.D., P.E.**
Virginia Tech Transportation Institute
3500 Transportation Research Plaza
Blacksburg, VA 24061
Phone: (540) 231-2673
Fax: (540) 231-1555
jdu@vtti.vt.edu

**Hesham A. Rakha, Ph.D., P.Eng. (Corresponding author)**
Charles E. Via, Jr. Department of Civil and Environmental Engineering
Virginia Tech Transportation Institute
Virginia Polytechnic Institute and State University
3500 Transportation Research Plaza
Blacksburg, VA 24061
Phone: (540) 231-1505
Fax: (540) 231-1555
hrakha@vt.edu

**Helena Breuer**
Virginia Tech Transportation Institute
3500 Transportation Research Plaza
Blacksburg, VA 24061
ahmed_elbery@mena.vt.edu

## ABSTRACT

With the rapid increase in ride-hailing use, a need to better understand and regulate the industry arises. Conflicting results have been published by researchers and policy makers regarding ride-hailing's impact on congestion, public transit, and other aspects of traffic systems. One of the obstacles to investing in ride-hailing is the lack of granular operational data over a relatively long period of time. To build an efficient system that can coordinate ride-hailing and other travel modes to better serve travel needs and minimize negative impacts on the transportation system requires enough ride-hailing trip-by-trip data and an understanding of ride-hailing trip patterns. Such patterns include the times that travelers use ride-hailing, where they are traveling from and to, how weekend and weekday ride-hailing trips differ, etc. This paper analyzes a year's worth of ride-hailing trip data from the Greater Chicago Area, which included detailed time, date, trip length, origin, and destination information to study the ride-hailing trip patterns. More than 104 million trips were analyzed. For trip rates, the results show that the total number of trips remained stable over the year, with pooled trips steadily decreasing from 20% to 9%. People tend to use RH more on weekends compared to weekdays. Specifically, weekend RH trip counts (per day) are, on average, 20% higher than weekday trip counts. The results of this work will help policy makers and transportation administrators better understand the nature of ride-hailing trips, which in turn allows for the design of a better regulation and guidance system for the ride-hailing industry.

## INTRODUCTION

Ride-hailing (RH) is defined as an on-demand, app-based, real-time service that provides customers with door-to-door transportation. First introduced to the market by Uber in 2008, RH has been a transportation option in the U.S. for more than a decade now. RH ridership at least tripled in the 3 years from 2016 to 2019, with active Uber users reaching 111 million during the last quarter of 2019 [1]. As of January 2019, 36% of U.S. adults had either used or were currently using RH services [2]. With such a dramatic increase in RH trips, the impacts of this relatively new industry on the transportation network need to be carefully studied, and corresponding policies should be designed to regulate its operation.

There have been multiple studies exploring RH trip patterns, impacts, correlation with other modes, optimization of ridesharing paring methodologies, etc. For example, one consideration is the overall change in vehicle miles traveled (VMT) that RH may bring to the transportation system. Some research has concluded that pooling multiple travelers together will help decrease the overall number of cars and the resulting VMT through a pairing and optimization process [3-16]. Other studies, however, have concluded that ridesharing increases the overall mileage traveled by personal vehicles. This increase has been attributed to multiple causes: detour driving for picking up and dropping off passengers, induced travel by individuals who are not be able to drive if no ridesharing service is available, diverted travelers from public transit, etc. [17-25]. Another conflicting argument comes from the relationship of RH with public transit [13]. Some studies believe that RH diverts travelers away from public transit, finding that after the introduction of RH services into the market, transit ridership began to decrease rapidly [15, 17, 23]. Others have found that RH contributes to a better use of the public transit system because it offers first-and last-mile connectivity with public transit or complements public transit when transit runs less frequently or is unavailable [16, 26]. There are also many studies concentrating on optimization methods or pairing modeling that help RH maximize the serving scope with minimum negative impact on congestion [8, 27-32]. Finally, a number of studies have concentrated on RH pricing strategies and revenues [33, 34].

The compound effect of RH on the transportation system will never be a simple question and should be investigated from multiple perspectives. For example, one of the key elements in determining RH's impact on VMT is calculating which will become the deciding factor: the number of passengers each RH trip can serve or the extra mileage incurred in an RH trip to pick up all passengers. To decide if RH benefits or disbenefits public transit, in-depth examination is needed to comprehensively investigate the time, origin, and destination of RH trips and the corresponding available public transit trip that may be used to replace an RH trip. As shown in Figure 1, RH interacts with mode shift, traveler matching, and routing, and also induces traffic demand. To evaluate the impacts of RH and optimize the efficiency of this trave l mode, a systematic modeling framework that incorporates all the factors is needed. Therefore, as a prerequisite to designing an effective system to coordinate RH with the existing transportation system to enhance RH's efficiency, we need to start by examining the patterns of RH trips and investigating their temporal and spatial distribution. To do so requires data at a high granularity over a long enough time period. Desirable detailed RH trip information to conduct a comprehensive and systematic analysis would include trip purpose, departure time, origin, and destination. Other factors that need to be considered include trip length, tolerance for (absolute) pick-up/drop-off delay or distance, tolerance for pick-up/drop-off delay with respect to the trip length or time, and tolerance for deviation from preferred departure time. However, since RH is a relatively new traffic mode and the market share has increased so rapidly, there have been limited studies that have collected such large-scale data. In addition, RH companies are typically reluctant to share their data due to privacy issues. Existing literature related to RH typically use the following major data sources: stated or revealed RH survey data, National Household Travel Survey data, large scale taxi data (references), local household travel data, census data, and/or simulation [29, 31, 35-46].

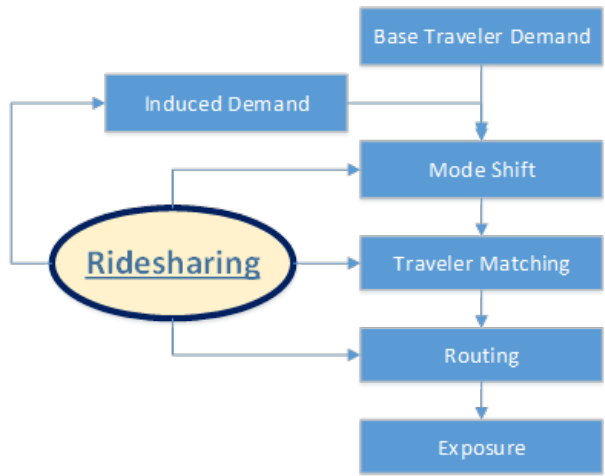

**Figure 1 Ridesharing interacting with other factors**

It is our belief that any effort to better regulate and improve the operation of the RH industry should be based on a comprehensive understanding of the pattens and variation of RH trips. Until there is collective concurrence about the effect of RH, designers, planners, and politicians cannot make sound decisions. The goal of this study, therefore, was to obtain a large-scale RH database, examine the temporal fluctuation of trips, investigate the patterns of single passenger RH trips (STs) as well as pooled RH trips (PTs), and make recommendations for better administration and regulation polices for the RH industry.

**DATASET ACQUISITION**

An ideal dataset for this study should include detailed trip-by-trip information, including pick up/drop off locations, departure time, number of passengers (to determine if the trip is a PT or an ST), trip purpose, trip distance and time, passenger demographic data, etc. However, to our knowledge, there is no such complete dataset available. First, RH trip datasets are scarce due to privacy issues and proprietary information concerns at RH companies. Secondly, RH is a relatively new industry. Finally, collecting and cleaning such a complete dataset requires some time. Despite these obstacles, we were able to find a close, if not perfectly matched, candidate dataset for our research.

In November 2018, the City of Chicago began collecting trip data from transportation network companies such as Uber and Lyft. The data are published to their open data portal (Chicago Data Portal [47]) and are available to the public. Three datasets are available: (1) trips, (2) drivers, and (3) vehicles. The trips database has 21 fields. Note: To protect traveler privacy, the pickup or dropoff locations for all trips are set to the centroid of the census tract where the trip origin and destination was located.

- Trip ID

- Pick-up: time, date, location, community area
- Drop-off: time, date, location, community area
- Trip duration (seconds) and distance
- Fare (rounded to the nearest $0.50)
- Tip (rounded to the nearest dollar)
- Additional charges
- Total trip cost
- Shared trip (true/false)
- Trips pooled
- Centroid pick-up: latitude, longitude
- Centroid drop-off: latitude, longitude

The trips dataset from Chicago is of particular interest because of the granularity of the data. It is the most detailed RH trip database that we are aware of and provides trip information detailed enough to serve the research purposes of this paper.

## DATA ANALYSIS

The data from January 1, 2019 to December 31, 2019 were downloaded for use in this paper. Table 1 summarizes the RH trips. The total number of RH trips fluctuated slightly over the year; more trips occurred during the spring, with a light dip in the summer and winter. There was a sharp decreasing trend in the percentage RH PTs, from 20% in January to 9% in December.

**TABLE** 1 **Total Number of Ridesharing Trips by Month (millions)**

|        | Total | Pooled (%)     | Month | Total | Pooled (%)     | Month | Total | Pooled (%)   |
|--------|-------|----------------|-------|-------|----------------|-------|-------|--------------|
| **Jan.** | 8.303 | 1.656 (20%) | **May** | 9.098 | 1.461 (16%) | **Sept.** | 8.332 | 0.831 (10%) |
| **Feb.** | 8.337 | 1.669 (20%) | **June** | 8.912 | 1.218 (14%) | **Oct.** | 8.871 | 0.785 (9%) |
| **March** | 9.664 | 1.827 (19%) | **July** | 8.562 | 0.976 (11%) | **Nov.** | 8.691 | 0.766 (9%) |
| **April** | 8.650 | 1.527 (18%) | **Aug.** | 8.807 | 0.921 (10%) | **Dec.** | 8.634 | 0.763 (9%) |

More RH was used on weekends compared to weekdays. On average, there were 24% more RH trips per day during weekends than during the weekdays. The rate of PTs, however, was higher on weekdays. The percentage of PTs ranged from 18.7% to 8.4% for weekend trips, while the percentage of PTs for weekdays ranged from 21.4% to 8.9%. The average percentage of PTs on weekdays was 4.5% higher than on weekends (Table 2). This pattern repeated itself across different months. The only exceptions occurred during holidays, especially when the holiday connected with a weekend. We found that travelers used RH much less during holiday seasons. Figure 2 and Figure 3 show the number of trips in December and May.

**TABLE 2 Trips by Weekend and Weekdays (per day, thousands)**

| Month | Trips/Weekend Day (% Pooled) | Trips/ Weekday (% Pooled) | Month | Trips/Weekend Day (% Pooled) | Trips/ Weekday (% Pooled) |
|---|---|---|---|---|---|
| **January** | 299 (18.3%) | 242 (21.6%) | **July** | 315 (10.9%) | 248 (11.8%) |
| **February** | 336 (18.7%) | 269 (21.3%) | **August** | 323 (10%) | 252 (11%) |
| **March** | 349 (17.4%) | 276 (20.7%) | **September** | 315 (9.5%) | 245 (10.5%) |
| **April** | 324 (16.2%) | 264 (18.8%) | **October** | 333 (8.5%) | 256 (9.1%) |
| **May** | 326 (15.3%) | 270 (16.7%) | **November** | 315 (8.4%) | 265 (9.4%) |
| **June** | 332 (12.9%) | 266 (14.5%) | **December** | 298 (8.4%) | 262 (9.2%) |

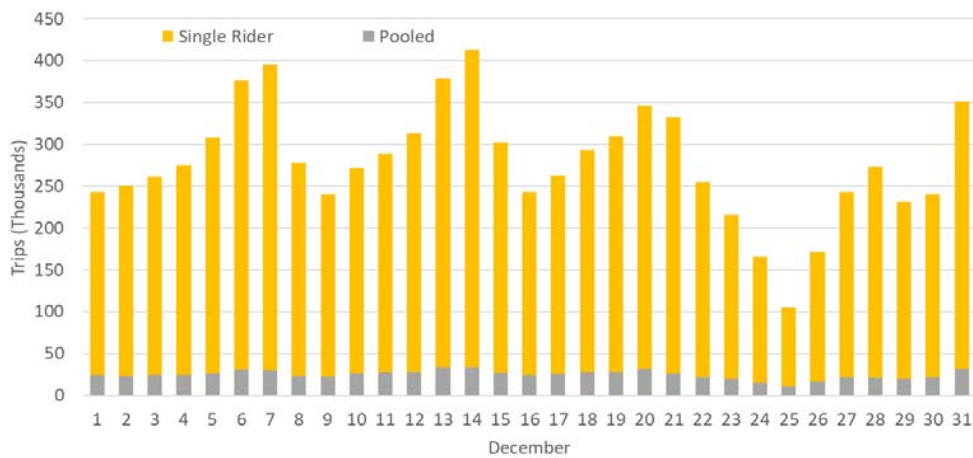

**Figure 2 Typical day-to-day variation in December**

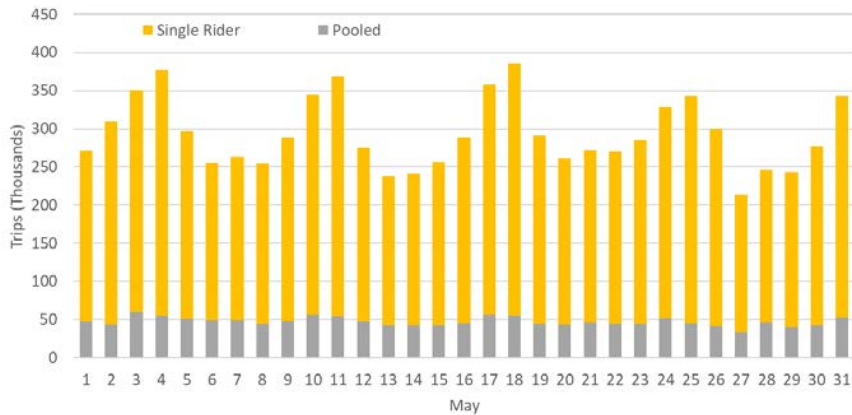

**Figure 3 Typical day-to-day variation in May**

Hourly variations were different for weekdays and weekends. Weekday trips had obvious morning and afternoon peaks, concurring with the typical morning and afternoon commuting peaks. Weekend RH trips peaked at about 6:00–7:00 p.m. and continued at a high level of demand until after midnight. The percentage of PTs for weekends was relatively higher during the day compared to the night. Meanwhile the PTs for weekday trips peaked along with the peak hours. Figure 4 shows the hourly distribution of weekend versus weekday trips for the month of May. The labels in Figure 4 are the percentages of PTs. Hourly variations and PT percentages followed a similar pattern across different months. Weekday trips had a much high PT percentage during the evening peak hour through midnight while weekend trips had a relatively stable PT percentage (Figure 5).  As the data shows, the percentage of PTs decreased consistently over the months.

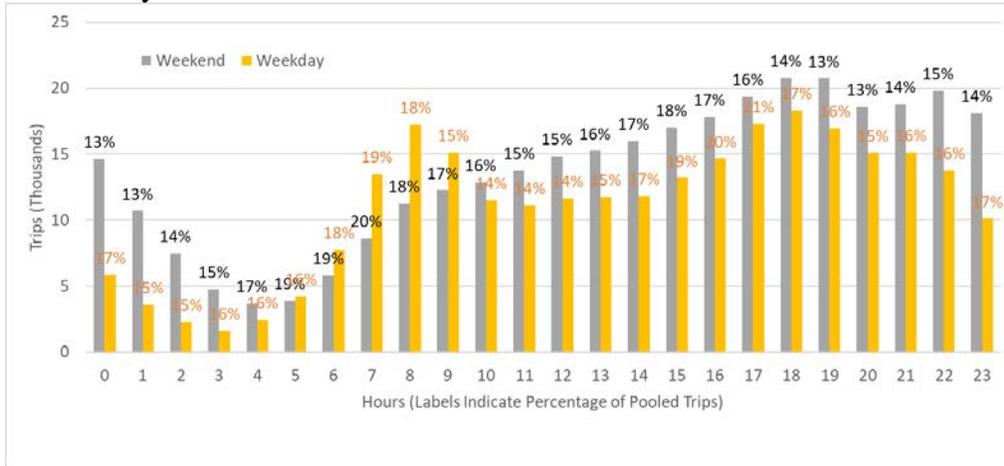

**Figure 4 hourly distribution of trips in May**

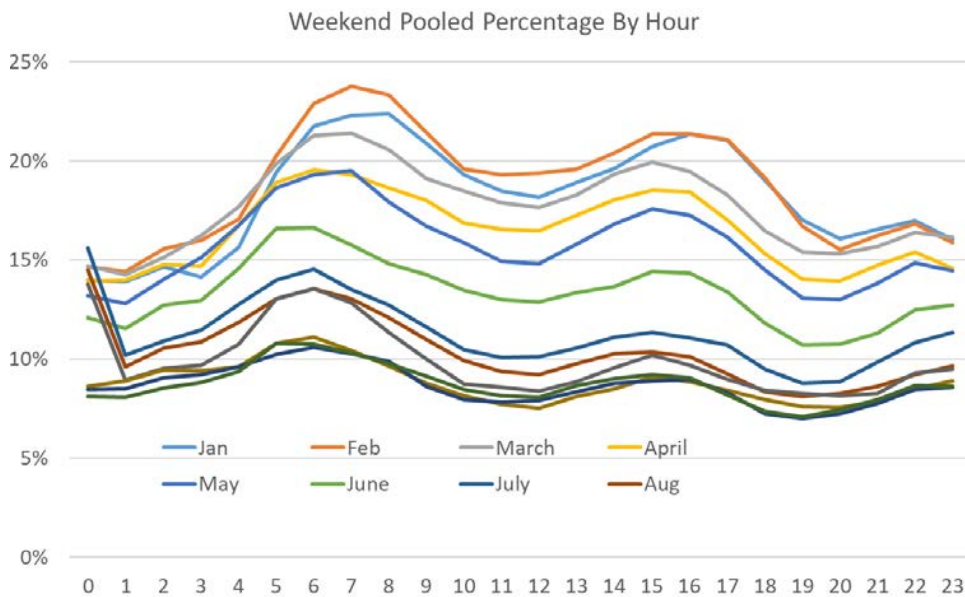

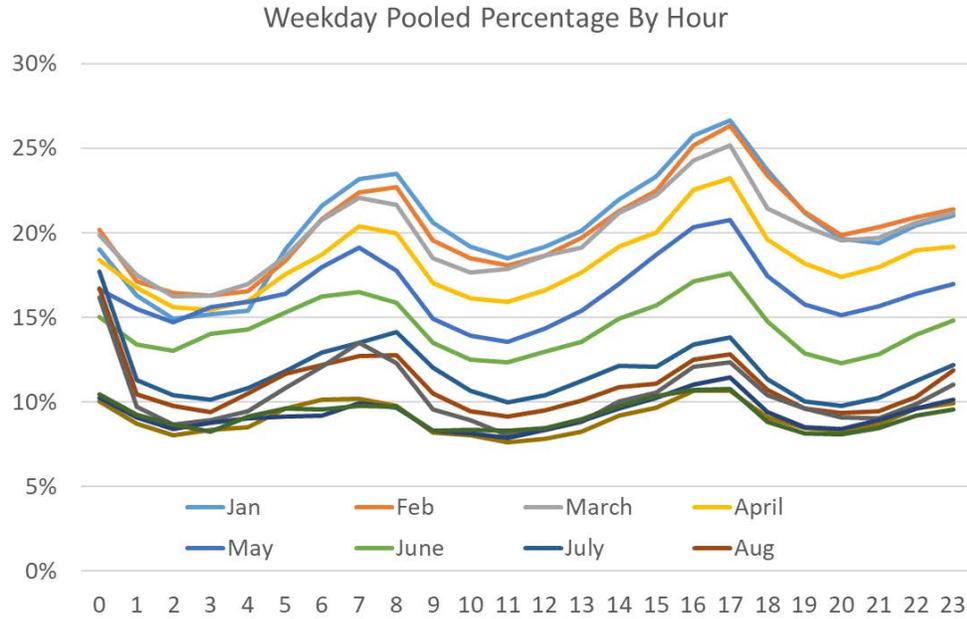

**Figure 5 PT percentages by weekends and weekdays**

For the lengths and travel times, as Figure 6 shows, more than 50% of the RH trips were shorter than 15 minutes and more than 60% were less than 5 miles. Travel times and distance distributions all follow a log-normal distribution, as shown the Q-Q plot in Figure 7. Therefore, the means and standard deviation (SD) for each group (PT versus ST, weekend versus weekday) can be compared directly. Table 3 lists the means and SDs for each trip group. Figure 8 illustrates the cost, travel distance, and time over 12 months. As expected, PTs were, on average, longer both time- and distance-wise than STs, while they cost much less for each passenger. In terms of travel time for STs, both weekdays and weekend trips were slightly longer during spring and summer (March to July) than during the rest of the year. PTs, however, were longer in winter months (October, November, December) for both weekdays and weekends. As the data shows, overall, the means and SD for all groups were relatively larger during spring to autumn. Winter typically had the smallest means in travel times, distances, and costs. The only exception were the PTs from October to December, when trips increased both in means and SDs. STs fluctuated the least in distances for both weekends and weekdays, though they did vary more in the spring in terms of cost and travel times. PTs had a larger mean in travel time and SDs during summer.

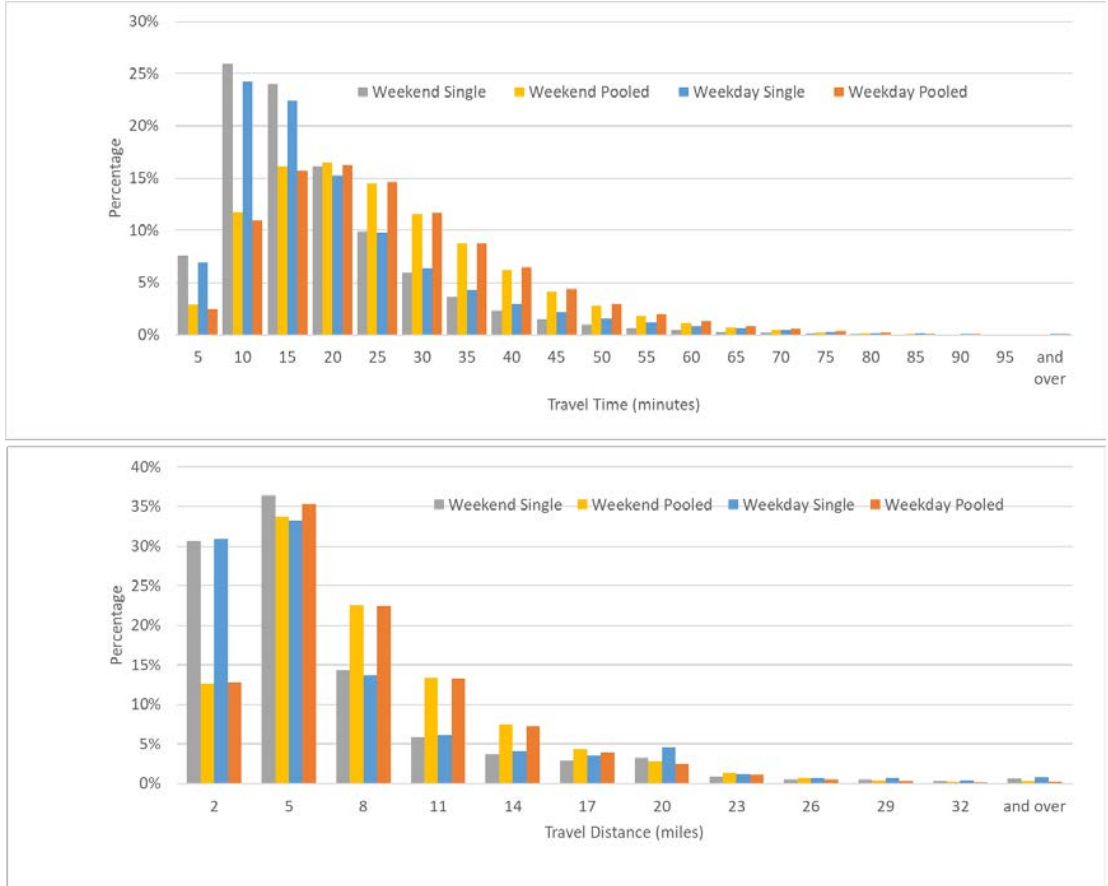

**Figure 6 Travel time and distance in May**

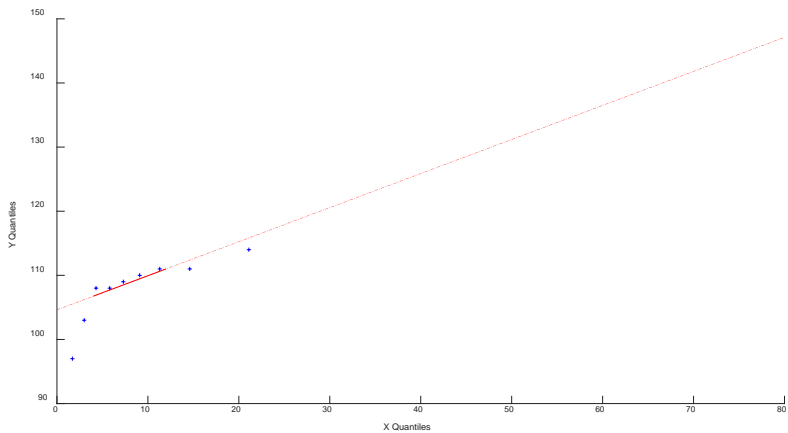

**Figure 7 QQ plot for travel time distribution against the log-normal distribution**

**TABLE** 3 **Means and Standard Deviation**

| | Jan | Feb | March | April | May | June | July | Aug | Sept | Oct | Nov | Dec |
|---|---|---|---|---|---|---|---|---|---|---|---|---|
| **WD-Pool Cost** | 3.68 (2.64) | 3.58 (2.64) | 3.61 (2.68) | 3.91 (2.99) | 4.09 (3.18) | 4.32 (3.24) | 4.68 (3.55) | 4.72 (3.59) | 4.76 (3.61) | 4.71 (3.2) | 4.79 (3.31) | 4.91 (3.23) |
| **WD -Pool Dist** | 5.97 (4.69) | 5.96 (4.72) | 6.13 (4.83) | 6.33 (4.96) | 6.6 (5.19) | 6.64 (5.14) | 6.67 (5.22) | 6.75 (5.3) | 6.81 (5.29) | 8.42 (6.52) | 8.65 (6.9) | 8.91 (6.68) |
| **WD -Pool Time** | 22.13 (12.65) | 22.45 (13.41) | 22.38 (12.98) | 23.08 (13.57) | 24.23 (14.33) | 24.18 (14.01) | 23.91 (13.79) | 23.69 (13.54) | 24.44 (14.13) | 24.02 (13.99) | 24.31 (14.49) | 23.58 (13.7) |
| **WD -Single Cost** | 14.21 (10.8) | 14.81 (11.21) | 14.9 (11.21) | 16.22 (12.81) | 16.82 (13.3) | 17.14 (13.61) | 15.49 (11.56) | 15.69 (11.52) | 16.17 (11.77) | 15.84 (11.48) | 15.67 (11.4) | 15.16 (10.78) |
| **WD -Single Dist** | 5.2 (6.28) | 5.41 (6.42) | 5.49 (6.44) | 5.71 (6.62) | 5.82 (6.71) | 5.87 (6.81) | 5.81 (6.81) | 5.97 (6.91) | 6.09 (6.88) | 5.97 (6.79) | 5.76 (6.63) | 5.64 (6.5) |
| **WD -Single Time** | 15.32 (11.15) | 16.25 (12.04) | 16.32 (12.01) | 17.18 (12.91) | 18.13 (13.96) | 18.35 (14.04) | 17.87 (13.23) | 17.84 (13.06) | 18.28 (13.45) | 17.85 (13.3) | 17.57 (13.22) | 16.65 (12) |
| **Wknd-Pool Cost** | 3.89 (2.76) | 3.8 (2.75) | 3.93 (2.87) | 4.16 (3.07) | 4.43 (3.45) | 4.79 (3.59) | 5.05 (3.75) | 5.12 (3.76) | 5.14 (3.94) | 5.04 (3.4) | 4.93 (3.14) | 5.13 (3.26) |
| **Wknd-Pool Dist** | 6.24 (4.97) | 6.1 (4.87) | 6.34 (5.03) | 6.54 (5.14) | 6.85 (5.44) | 6.89 (5.39) | 6.89 (5.47) | 7.05 (5.57) | 7.06 (5.59) | 8.93 (6.86) | 9.44 (7.21) | 9.6 (6.98) |
| **Wknd-Pool Time** | 21.5 (12.42) | 21.79 (12.88) | 22.02 (12.87) | 22.01 (12.73) | 23.44 (13.9) | 23.47 (13.64) | 22.81 (13.13) | 23.16 (13.26) | 23.27 (13.47) | 23.53 (13.91) | 23.28 (13.44) | 23.24 (13.24) |
| **Wknd-Single Cost** | 13.55 (9.66) | 13.77 (9.6) | 14.24 (10.01) | 14.98 (10.92) | 16.25 (12.09) | 16.68 (12.39) | 14.77 (10.16) | 14.82 (10.07) | 15.06 (10.49) | 15.08 (10.34) | 14.48 (9.64) | 14.69 (9.94) |
| **Wknd-Single Dist** | 4.89 (5.89) | 4.81 (5.72) | 5.01 (5.94) | 5.14 (6.03) | 5.3 (6.15) | 5.36 (6.19) | 5.36 (6.21) | 5.48 (6.33) | 5.56 (6.43) | 5.51 (6.33) | 5.39 (6.16) | 5.46 (6.24) |
| **Wknd-Single Time** | 14.29 (10.09) | 14.71 (10.37) | 15.25 (10.69) | 15.31 (10.92) | 16.29 (11.8) | 16.67 (11.85) | 16.35 (11.33) | 16.59 (11.44) | 16.52 (11.66) | 16.64 (12.1) | 15.8 (10.9) | 15.82 (10.87) |

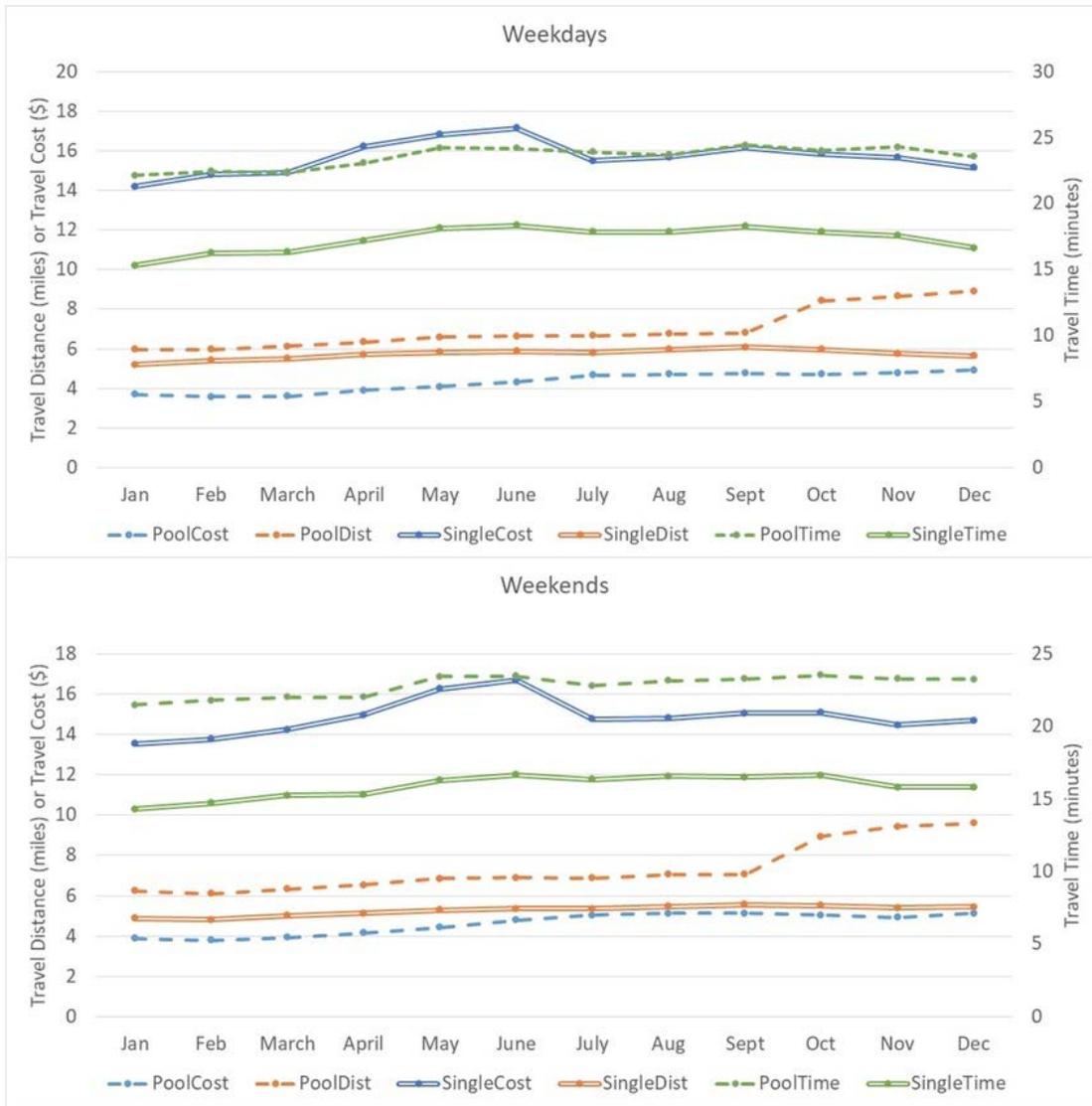

**Figure** 8 **Travel cost (per person) time and distance distribution**

Table 4 lists the results of two sample T-test results for weekday vs. weekend trips. As the table shows, PTs were not significantly different in cost and travel distance for the two groups. However, STs were significantly different between the two groups, with weekday trips being longer in travel distance or time and more expensive than weekend trips.

**TABLE 4 Two-Sample T-Test Weekday Vs. Weekend**

|  | *T  (P-value)* | *Difference in Means* |
|---|---|---|
| *Pooled Cost* | -1.417 (0.17) | [-0.749, 0.141] |
| *Pooled Dist* | -0.721 (0.478) | [-1.321, 0.639] |
| *Pooled Time* | 2.312 ($< 0.05$) | [0.076, 1.404] |
| *Single Cost* | 2.288 ($< 0.05$) | [0.076, 1.549] |
| *Single Dist* | 4.39 ($< 0.05$) | [0.241, 0.671] |
| *Single Time* | 4 ($< 0.05$) | [0.697, 2.198] |

## CONCLUSION AND DISCUSSION

This paper analyzed a large trip-based ride-hailing dataset collected in Chicago covering the whole year of 2019. The goal of this research was to conduct an in-depth data exploration of RH trips in terms of trip attributes and temporal variations. The results of the study and discussions are summarized as follows.

For trip rates, the results show that the total number of trips remained stable over the year, with pooled trips steadily decreasing from 20% to 9%. This appears to indicate that either travelers prefer to travel alone after getting familiar with the RH travel mode or that travelers were not satisfied with the pooled trip experience. One positive contribution of RH to the transportation system is its capability of combining multiple trips such that the total mileage traveled can be decreased. Our results showed that unless effective polices are enforced in the future to encourage this positive contribution by urging passengers to travel together, RH will generate extra mileage within the traffic network.

People tend to use RH more on weekends than weekdays. Weekend RH trip counts (per day) are, on average, 20% higher than weekday trip counts. If a holiday is connected with a weekend, however, that trend is no longer true. For the hourly distribution, weekday trips align with typical commuting morning and evening peaks. Weekend trips occur more after 5:00 p.m. through midnight. Among these trips, the rate of pooled trips is consistently higher during the weekdays, with three obvious peaks—(1, 2) morning and evening commuting peaks and (3) midnight—while weekend PT percentages have a relatively smaller fluctuation over the 24 hours. Since no trip purpose information is available in the database, we can only infer from these observations, based on the trip departure times, that travelers tend to use RH more for leisure trips during weekends. Meanwhile, when travelers use RH on weekdays, they are more likely to car-pool with others for commuting and nighttime leisure trips.

The time and distance distribution for either weekday or weekend and PTs or STs is similar. RH trips are concentrated in the range of shorter than 15 minutes and less than 5 miles, thus indicating  that travelers use RH for short trips. Comparing between weekend and weekday trips, we observe that travel time and distance for STs are significantly different for weekdays and weekends. Specifically, weekday trips are longer than weekend trips in both distance and time, and are therefore more expensive.

Our conclusions are compatible with previous studies using stated-preference surveys in that travelers tend to use RH for short leisure trips during weekends and for work trips during weekdays. PTs only account for a small percentage of all RH trips. Further, weekday and weekend trips are statistically different from each other. These conclusions indicate that RH may not be a solution for decreasing traffic congestion and VMT if no other effective regulations are imposed to encourage and reward pooled trips. Meanwhile, polices should be formulated to help the RH industry coordinate with other travel modes to better serve travel needs and minimize the negative impacts of this newly developed travel mode.

## ACKNOWLEDGMENTS

This paper is sponsored by The Urban Mobility & Equity Center (UMEC).

## AUTHOR CONTRIBUTIONS

The authors confirm contribution to the paper as follows: study conception and design, Jianhe Du and Hesham Rakha; data collection, Jianhe Du and Helena Breuer; data analysis and interpretation of results, Jianhe Du and Hesham Rakha; draft manuscript preparation, Jianhe Du, Hesham Rakha, and Helena Breuer. All authors reviewed the results and approved the final version of the manuscript.